\documentclass[aps,prb,showpacs,groupedaddress,twocolumn]{revtex4}
\usepackage{amsmath,amssymb,bm,units}
\usepackage{graphicx}

\DeclareMathOperator{\artanh}{artanh}
\DeclareMathOperator{\sech}{sech}

\newcommand{\DLT}[1][ ]{\mathcal{D}_{L/T#1}}
\newcommand{\DT}[1][ ]{\mathcal{D}_{T#1}}
\newcommand{\DL}[1][ ]{\mathcal{D}_{L#1}}
\newcommand{\T}[1][ ]{\mathcal{T}_{#1}}

\newcommand{\jS}[1][ ]{j_{S#1}}

\newcommand{\abs}[1]{\left\lvert{#1}\right\rvert}

\newcommand{\dd}[1]{\mathrm{d}#1\,}

\newcommand{\mat}[1]{\begin{pmatrix}#1\end{pmatrix}}
\newcommand{\uvec}[1]{\hat{\mathbf{#1}}}

\begin{document}

\title{Peltier effect in Andreev interferometers}
\date{\today}

\author{Pauli Virtanen}
\email[]{Pauli.Virtanen@tkk.fi}
\author{Tero T. Heikkil\"a}
\email[]{Tero.T.Heikkila@tkk.fi}
\affiliation{Low Temperature Laboratory, Helsinki University of
  Technology, P.O. Box 2200 FIN-02015 TKK, Finland.}

\begin{abstract}
  The superconducting proximity effect is known to modify transport
  properties of hybrid normal--superconducting structures.  In
  addition to changing electrical and thermal transport separately, it
  alters the thermoelectric effects.  Changes to one off-diagonal
  element $L_{12}$ of the thermoelectric matrix $L$ have previously
  been studied via the thermopower, but the remaining coefficient
  $L_{21}$ which is responsible for the Peltier effect has received
  less attention.  We discuss symmetry relations between $L_{21}$ and
  $L_{12}$ in addition to the Onsager reciprocity, and calculate
  Peltier coefficients for a specific structure.  Similarly as for the
  thermopower, for finite phase differences of the superconducting
  order parameter, the proximity effect creates a Peltier effect
  significantly larger than the one present in purely normal-metal
  structures. This results from the fact that a nonequilibrium
  supercurrent carries energy.
\end{abstract}


\maketitle

In large metallic structures, linear-response transport can be
described using the thermoelectric matrix $L$ that relates charge and
energy currents to temperature and potential biases.  \cite{callen48}
The off-diagonal coefficients describe coupling between heat and
charge currents, and indicate the magnitude of the thermopower and the
Peltier effect. In many cases, these coefficients are coupled by
Onsager's reciprocal relation $L_{\alpha\beta}(B)=L_{\beta\alpha}(-B)$
under the reversal of the magnetic field $B$.
\cite{onsager31,callen48}

\begin{figure}
  \includegraphics{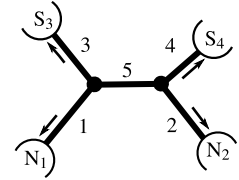}
  \caption{
    \label{fig:4probe}
    Example of a 4-probe structure considered in the text: 5
    normal-metal wires connected to each other and to 4 terminals, of
    which two are superconducting (S) and two normal (N).  We take the
    lengths $l$, cross-sectional areas $A$, and conductivities
    $\sigma$ of the wires to be $l/l_0=(1.5,1,1.2,1,0.8)$ and
    $A\sigma/A_0\sigma_0=(0.8,1,0.8,1,1)$. Here $l_0$, $A_0$ and
    $\sigma_0$ are some characteristic values, controlling the
    energy scale $E_T=\hbar D/(l_3+l_4+l_5)^2$ of the proximity
    effect. The system is chosen so as to bring out effects that
    depend on the magnitude of geometrical asymmetry. In the numerics,
    all wires are assumed to be quasi-one-dimensional, $l\gg\sqrt{A}$.
  }
\end{figure}

In hybrid normal--superconducting systems (see Fig.~\ref{fig:4probe})
the Cooper pair amplitude penetrates to the normal-metal parts.  This
makes the linear-response coefficients $L$ different from their
normal-state values, and allows supercurrents $I_{S,\mathrm{eq}}$ to
flow through the normal metal even at equilibrium.  The charge and
energy (entropy) currents $I_c^i$ and $I_E^i$ entering different
terminals can in linear response be written as
\begin{align}\label{eq:linearresponseNS}
  \mat{ I_c^{i} - I_{S,\mathrm{eq}}^{i} \\ I_E^{i} }
  = 
  \sum_{j\in\mathrm{terminals}}
  \mat{ L_{11}^{ij} & L_{12}^{ij} \\
        L_{21}^{ij} & L_{22}^{ij} }
  \mat{\Delta V_j \\ \Delta T_j / \bar{T}}
  \,,
\end{align}
in terms of the biases $\Delta{}V_j=V_j-\bar{V}$,
$\Delta{}T_j=T_j-\bar{T}$ and the modified response coefficients $L$.
The proximity-induced changes in the conductance $L_{11}^{ij}$,
\cite{charlat96,claughton96} thermal conductance $L_{22}^{ij}/\bar{T}$ (for
$L_{12}^{ij}=L_{21}^{ij}=0$) \cite{claughton96,bezuglyi03,jiang04b,jiang05} and
thermopower $-L_{12}^{ij}/(\bar{T}L_{11}^{ij})$
\cite{claughton96,eom98,seviour00,kogan02,parsons03,parsons03b,virtanen04,virtanen04b,jiang05b,volkov05,giazotto2006-opportunities}
have recently been investigated both experimentally and theoretically.
Behavior of the remaining off-diagonal coefficient $L_{21}^{ij}$ has
previously been discussed in Ref.~\onlinecite{claughton96} using
scattering theory, but the simulations were restricted to small
structures --- making the contribution from electron-hole asymmetry
very large.

In this article, we note that within reasonable approximations, in
diffusive superconducting heterostructures
Eq.~\eqref{eq:linearresponseNS} can be generalized to the non-linear
regime by defining an energy-dependent thermoelectric matrix
$\tilde{L}_{\alpha\beta}^{ij}(E)$. We show that this quantity
satisfies an Onsager reciprocal relation
$\tilde{L}_{\alpha\beta}^{ij}(E,B)=\tilde{L}_{\beta\alpha}^{ji}(E,-B)$
under the reversal of the magnetic field $B$ and the phase
$\arg\Delta$ of the superconducting order parameter, whenever $i$ and
$j$ refer to normal terminals. We also show how the proximity effect
modifies $L_{21}$ giving rise to a large Peltier effect, \cite{callen48} and
discuss how it could be experimentally detected.

Qualitatively, one can understand the origin of proximity-induced
thermoelectric effects by noting that charge current consists of a
quasiparticle component and a supercurrent component.  That the latter
is strongly temperature dependent in proximity structures then leads
to a finite $L_{12}$ coefficient, \cite{seviour00,virtanen04} via a
mechanism analogous to charge imbalance generation in superconductors.
\cite{schmid79,pethick79} Assuming Onsager symmetry, one would also
expect that $L_{21}$ is finite. The actual form of the coupling can be
seen by inspecting the quasiclassical transport equations
(Eqs.~\eqref{eq:kin} below), or by studying their near-equilibrium
approximation in a diffusive normal metal under the influence of a
weak proximity effect (see for example
Ref.~\onlinecite{virtanen04b}):
\begin{subequations}
\label{eq:linearizedkin}
\begin{align}
  \nabla\cdot J_c &= 0 \,, \quad \nabla\cdot J_E = 0 \,,
  \\
  \label{eq:linearizedjc}
  J_c &= -\tilde{\sigma} \nabla\delta V + \tilde{\cal T}\nabla\delta T
          + \frac{\partial J_{S,\rm eq}}{\partial T} \delta T 
          + J_{S,\rm eq}
  \,, 
  \\
  \label{eq:linearizedje}
  J_E &= -\tilde{\sigma}_{\rm th} \nabla\delta T - \bar{T} \tilde{\cal T}\nabla\delta V + \bar{T} \frac{\partial J_{S,\rm eq}}{\partial T} \delta V \,.
\end{align}
\end{subequations}
Here, $\delta V$ and $\delta T$ are deviations of the (effective)
local potential and temperature from equilibrium, and $\bar{T}$ is the
ambient temperature.  The first terms in charge and energy current
densities $J_c$, $J_E$ can be considered the quasiparticle current and
the rest the (non-equilibrium) supercurrent; $\tilde{\sigma}$ and
$\tilde{\sigma}_{\rm th}$ are the proximity-modified charge and
thermal conductivities, $J_{S,\rm eq}$ is the equilibrium supercurrent
density, and $\tilde{\cal T}$ is a small factor associated with
non-equilibrium supercurrent. Although Eqs.~\eqref{eq:linearizedkin}
are not of the usual form of normal-state transport equations,
\cite{callen48} one can see that a variation $\delta T$ generates a
change in the charge current, and that non-equilibrium
($\delta{}V\ne0$) supercurrent carries energy current. The
corresponding response coefficients in Eqs.~\eqref{eq:linearizedjc}
and~\eqref{eq:linearizedje} are not independent, which is a signature
of the Onsager symmetry. Comparing the magnitude of the coefficients,
it turns out that at low temperatures a large part of the
thermoelectric coupling indeed arises from the temperature-dependence
of $J_{S,\rm{}eq}$. At high temperatures where it vanishes
exponentially, other sources become more important.
\cite{volkov05,kogan02,virtanen04b}

However, validity of Eqs.~\eqref{eq:linearizedkin} is somewhat
restricted, since they are correct only in the linear response and to
the first order in the proximity corrections, additionally assuming
that the energy gap $\abs{\Delta}$ of the nearby superconductors
satisfies $k_B\bar{T}\ll\abs{\Delta}$. For quantitative calculations of
the multiterminal transport coefficients, and to evaluate the
proximity-corrected coefficients in Eq.~\eqref{eq:linearizedkin}, we
start from the full non-equilibrium formalism.

The superconducting proximity effect can be described using the
quasiclassical BCS--Gor'kov theory.  \cite{kopnin01,belzig99} Here, we
concentrate on diffusive normal-metal structures that are connected to
superconducting and normal terminals, and neglect any inelastic
scattering. The model then reduces to the Usadel equations,
\cite{usadel70,belzig99} whose first part, the spectral equations, can
in this case be written as
\begin{subequations}
  \label{eq:spectral}
  \begin{gather}
    D \nabla^2 \theta = -2i(E + i 0^+) \sinh\theta
    + \frac{v_S^2}{2D}\sinh(2\theta) \,,
    \label{eq:spectral1}
    \\
    \nabla\cdot(-v_S\sinh^2\theta) = 0 \,,
    \quad v_S \equiv D(\nabla\chi - 2e{\bf A}/\hbar)
    \,.
    \label{eq:spectral2}
  \end{gather}
\end{subequations}
They describe the penetration of the superconducting pair amplitude
$F=e^{i\chi}\sinh\theta$ into the normal metal. We denote the
diffusion constant of the metal here by $D$, and the magnetic vector
potential by ${\bf A}$.  At clean contacts to bulk superconductors, the
pairing angle is $\theta=\artanh(\abs{\Delta}/E)$ and the phase
$\chi=\arg\Delta$, where $\Delta$ is the superconducting order
parameter. Transport properties are in turn determined by kinetic
Boltzmann-like equations
\begin{subequations}
  \label{eq:kin}
  \begin{gather}
    D\nabla\cdot \hat\Gamma_T f = {\cal R} f_T + D(\nabla\cdot{}j_S) f_L,
    \;
    D\nabla\cdot \hat\Gamma_L f = 0,
    \\
    \hat\Gamma_T f \equiv \DT\nabla f_T + \T\nabla f_L + \jS f_L
    \,,
    \\
    \hat\Gamma_L f \equiv \DL\nabla f_L - \T\nabla f_T + \jS f_T
    \,,
  \end{gather}
\end{subequations}
that describe the behavior of the antisymmetric and symmetric parts
$f_L(E)\equiv{}f(\mu_S-E)-f(\mu_S+E)$ and
$f_T(E)\equiv{}1-f(\mu_S-E)-f(\mu_S+E)$ of the electron distribution
function. They are defined with respect to the potential of the
superconductors, chosen below as $\mu_S=0$.  The spectral supercurrent
$\jS$, the diffusion coefficients $\DL$, $\DT$, $\T$, and the
condensate sink term ${\cal R}$ are functionals of $\theta$ and
$\chi$, having the symmetries $\DLT(\chi)=\DLT(-\chi)$,
$\T(\chi)=-\T(-\chi)$, $j_S(\chi)=-j_S(-\chi)$, ${\cal R}(\chi)={\cal
  R}(-\chi)$. \cite{belzig99,virtanen04b} In normal metals,
$\nabla\cdot{}j_S={\cal R}=0$. Observable current
densities are finally related to the spectral currents
$\hat\Gamma_{L/T} f$ through
\begin{align}
  \label{eq:currents}
  J_c = -\frac{\sigma}{2\abs{e}}\int_{-\infty}^{\infty}\dd{E} \hat\Gamma_T f \,, \;
  J_E = \frac{\sigma}{2 e^2} \int_{-\infty}^{\infty}\dd{E} E\,\hat\Gamma_L f \,.
\end{align}
and the heat current density is $J_Q = J_E - V J_c$ at the terminals. Below,
we also assume that all contacts to terminals are clean and of
negligible resistance: in this case all quantities are continuous at
the interfaces, except at superconductors for $E<\abs{\Delta}$ the
boundary condition for the kinetic L-mode is changed to
$\uvec{n}\cdot\hat{\Gamma}_Lf=0$, where $\uvec{n}$ is the normal to the
interface.

It is important to note that the last two terms in Eqs.~\eqref{eq:kin}
mix the L and T modes and cause thermoelectric effects: near
equilibrium, they lead to the coupling terms in
Eqs.~\eqref{eq:linearizedkin}. Away from linear response, a
non-equilibrium modification of the distribution function $f$ due to
the mixing \cite{heikkila03} has also been experimentally observed in
Ref.~\onlinecite{crosser06}.

The aim in the following is to calculate the thermoelectric
coefficients $L_{\alpha\beta}^{ij}$ starting from Eqs.~\eqref{eq:kin}.
However, as with the charge conductance, it is useful to first define
corresponding energy-dependent thermoelectric coefficients
$\tilde{L}_{\alpha\beta}^{ij}(E)$.  Since the kinetic equations are
linear, it is possible to write the currents entering different
terminals as
\begin{subequations}\label{eq:Gmatrix}
\begin{align}
  I_{c}^{i} &= \int_{-\infty}^\infty \dd{E}\,
             \sum_{\beta j} \tilde{L}_{T\beta}^{ij}(E) f_{\beta}^{j}(E)
  \,,
  \\
  I_{E}^{i} &= \int_{-\infty}^\infty \dd{E}\,E\,
             \sum_{\beta j} \tilde{L}_{L\beta}^{ij}(E) f_{\beta}^{j}(E)
  \,,
\end{align}
\end{subequations}
where $\beta\in\{T, L\}$, $j$ runs over all terminals, and
$f_\alpha^j$ is the $\alpha$-mode distribution function in terminal
$j$.  This spectral thermoelectric matrix
$\tilde{L}_{\alpha\beta}^{ij}(E)$ is the quasiclassical counterpart to
the $P$-matrix in Ref.~\onlinecite{claughton96}. More explicitly,
$\tilde{L}_{\alpha\beta}^{ij}(E)$ can be defined as the
$\alpha$-mode current seen in terminal $i$ that a unit excitation of
mode $\beta$ in terminal $j$ generates at energy $E$:
\begin{align}\label{eq:Ldefinition}
  \tilde{L}^{ij}_{\alpha\beta}(E) \equiv
  \int_{\mathcal{S}_i} \dd{\mathcal{S}} \,
  \uvec{n} \cdot \hat\Gamma_{\alpha} \psi^{j,\beta} 
  \,.
\end{align}
Here, $\mathcal{S}_i$ is the surface of terminal $i$ and $\uvec{n}$
the corresponding normal vector. The two-component function
$\psi^{j\beta}$ is assumed to satisfy the kinetic
equations~\eqref{eq:kin} with the electron distribution functions
$f_\alpha^i$ in terminals replaced by $\delta_{\alpha\beta}\delta_{ij}$.
The linear-response coefficients $L$ are
directly related to $\tilde{L}(E)$ via Eq.~\eqref{eq:Gmatrix}, for
example $L_{11}=\frac{1}{2k_BT}\int\dd{E}
\tilde{L}_{TT}(E)\sech^2(\frac{E}{2k_BT})$ and
$L_{21}=\frac{-1}{2k_BT}\int\dd{E}
E\,\tilde{L}_{LT}(E)\sech^2(\frac{E}{2k_BT})$.

The spectral thermoelectric matrix depends only on $\theta$ and
$\chi$, but not on the distribution functions at the terminals.
Knowing the energy dependence of this matrix, one can directly
evaluate currents also away from linear response, if changes in the
order parameter $\Delta$ and any inelastic scattering can be
neglected. The matrix $\tilde{L}_{\alpha\beta}^{ij}(E)$ one can
evaluate numerically once $\theta$ and $\chi$ have been solved, and it
offers a feasible way to find the response of the circuit to different
types of excitations in the terminals.

An Onsager reciprocal relation for $\tilde{L}_{\alpha\beta}^{ij}(E)$
follows from the fact that the differential operator
$\hat{\mathcal{L}}$ in Eqs.~\eqref{eq:kin}, $\hat{\mathcal{L}}f=0$,
has the property
\begin{align}
  \hat{\mathcal{L}}(B)^\dagger 
  &= (-\nabla)\cdot\mat{\DT & \T \\ -\T & \DL}^\dagger(-\nabla) 
    + (-\nabla)\cdot\mat{0 & j_S \\ j_S & 0}^\dagger
    \\\notag
    &\quad - \mat{ {\cal R} & -D(\nabla\cdot{}j_S) \\ 0 & 0 }
    \\\notag
  &= \hat{\mathcal{L}}(-B) \,,
\end{align}
due to the symmetries of the coefficients under reversal of the phases
$\chi$, $\arg\Delta$ and the magnetic field $B$. Below, whenever we
discuss reversal of $B$, also reversal of the phases is implied.
Integration by parts now shows that for any volume $\Omega$ and
two-component functions $\phi$, $\psi$, we can write
\begin{align}
  \int_{\Omega}\dd{\mathcal{V}}[\psi^\dagger \hat{\mathcal{L}} \phi - \phi^\dagger \hat{\mathcal{L}}^\dagger \psi] = \int_{\partial\Omega}\dd{\mathcal{S}} \uvec{n} \cdot J \,,
\end{align}
where $\partial\Omega$ is the boundary of $\Omega$.  For the
differential operator here, the flux
$J=\psi^\dagger\hat\Gamma(B)\phi-\phi^\dagger\hat\Gamma(-B)\psi-j_S\psi^\dagger\sigma_1\phi$, $\sigma_1$ being the first spin matrix.
Now, we choose $\Omega$ to be the whole conductor, with $\phi$ and
$\psi$ such that $\phi=\psi^{j,\beta}$ satisfies the conditions in the
calculation for $\tilde{L}_{\alpha\beta}^{ij}(E,B)$ and
$\psi=\psi^{i,\alpha}$ the conditions for
$\tilde{L}_{\beta\alpha}^{ji}(E,-B)$. 
When both $i$ and $j$ refer to normal terminals,
we then find
\begin{gather}\label{eq:implicitonsager}
  \begin{split}
  0
  &= \int_{\partial\Omega}\dd{\mathcal{S}}\uvec{n}\cdot J \\
  &= \int_{\mathcal{S}_i}\dd{\mathcal{S}} \uvec{n}\cdot\hat\Gamma_{\alpha}(B) \phi
  - \int_{\mathcal{S}_j}\dd{\mathcal{S}} \uvec{n}\cdot\hat\Gamma_{\beta}(-B) \psi
  \,,
  \end{split}
\end{gather}
using the boundary conditions imposed on $\phi$ and $\psi$, and the
fact that $j_S=0$ at normal terminals.
Comparison of Eqs.~\eqref{eq:implicitonsager}
and~\eqref{eq:Ldefinition} reveals a reciprocal relation
\begin{align}\label{eq:onsagersymmetry}
  \tilde{L}_{\alpha\beta}^{ij}(E,B)=\tilde{L}_{\beta\alpha}^{ji}(E,-B)
  \,.
\end{align}
This implies that phase differences in the order parameter will be
similar sources for quasiclassical Peltier and Thompson effects as
they are for the thermopower discussed in
Refs.~\onlinecite{seviour00,kogan02,virtanen04,virtanen04b,volkov05}.
Similar relations exist also in the scattering theory.  \cite{claughton96}

\begin{figure}
  \includegraphics{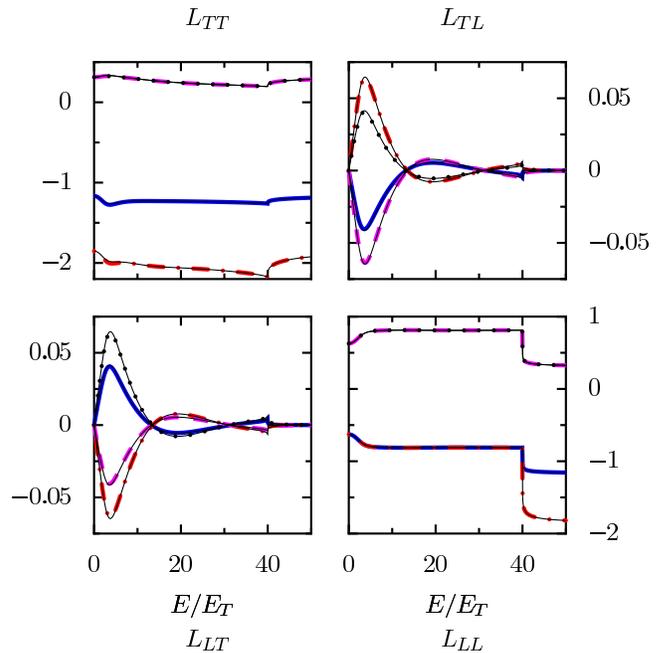}
  \caption{\label{fig:Lmatrix} {(Color online)} Elements
    $L^{ij}_{\alpha\beta}$ for $i,j=1,2$ and $\alpha,\beta=L,T$ for
    the asymmetric interferometer in Fig.~\ref{fig:4probe},
    in units of $1/R_2$.
    The order parameters in the superconducting terminals have
    $\arg\Delta_3 - \arg\Delta_4 = 0.54 \pi$ and
    $\abs{\Delta_3}=\abs{\Delta_4}=40E_T$.  Thick blue solid line is
    $L^{11}$, dash-dotted red line $L^{22}$, black dotted line $L^{12}$ and
    magenta dashed line $L^{21}$. Note the symmetry
    $\tilde{L}_{\alpha\beta}^{ij}(E)=(-1)^{1-\delta_{\alpha\beta}}\tilde{L}_{\beta\alpha}^{ji}(E)$.  
    Approximations found by
    solving Eqs.~\eqref{eq:kin} to first order in $j_S$ and
    $\mathcal{T}$ are shown as thin solid lines --- in general they
    are indistinguishable from the exact numerical results.  }
\end{figure}

The form of Eqs.~\eqref{eq:Gmatrix} also implies that $\tilde{L}(E,B)$ has
the symmetries
\begin{subequations}\label{eq:Lsymmetries}
\begin{gather}
  \sum_{j} \tilde{L}^{ij}_{TL}(E) = 0 \qquad\text{for normal terminal $i$,} \\
  \sum_{j} \tilde{L}^{ij}_{LL}(E) = 0 \,, \\
  \tilde{L}^{ij}_{\alpha\beta}(E,-B) 
  = (-1)^{1-\delta_{\alpha\beta}}\tilde{L}^{ij}_{\alpha\beta}(E,B)
  \label{eq:Lsymmetryqcl}
  \,,
\end{gather}
\end{subequations}
since the charge current to any normal terminal and the entropy
current to any terminal must vanish at equilibrium for all
temperatures. Equation~\eqref{eq:Lsymmetryqcl} follows essentially
from the electron-hole symmetry assumed in the quasiclassical theory,
leading to $\hat{\Gamma}_L{}f\mapsto{}\hat{\Gamma}_L{}f$,
$\hat{\Gamma}_T{}f\mapsto-\hat{\Gamma}_T{}f$ under the transformations
$B\mapsto-B$, $f_T\mapsto-f_T$. \cite{virtanen04b} This makes the
diagonal coefficients symmetric in $B$ and the off-diagonal ones
antisymmetric. However, there are some experimental results 
\cite{eom98,jiang05b}
where the latter symmetry does not hold. Such observations cannot be
explained with the quasiclassical theory applied here.

Consider now the application of the formulation above in the structure
in Fig.~\ref{fig:4probe}. We solve the spectral
equations~\eqref{eq:spectral} in this structure numerically and
calculate the spectral thermoelectric matrix from the solutions.
Behavior of the two coefficients important for thermoelectric effects,
spectral supercurrent $j_S$ and the coefficient $\mathcal{T}$, is
discussed for structures of this type for example in
Refs.~\onlinecite{heikkila02scdos,virtanen04b}.  Resulting elements of
$L^{ij}_{\alpha\beta}(E)$ are plotted as a function of $E$ in
Fig.~\ref{fig:Lmatrix} --- the energy scale is given by the Thouless
energy $E_T=\hbar D/(l_3+l_4+l_5)^2$. The diagonal elements
$\tilde{L}_{TT}(E)$ and $\tilde{L}_{LL}(E)$ are spectral charge and
energy conductances.  \cite{charlat96,bezuglyi03} At $E>\abs{\Delta}$,
energy current can enter also the superconductor, which is visible as
a rapid change in the $\tilde{L}_{LL}$-coefficient. The off-diagonal
coefficients qualitatively follow the energy dependence of the
spectral supercurrent $j_S$ which gives the most visible contribution.
Moreover, the elements of the matrix clearly exhibit the
symmetries~\eqref{eq:onsagersymmetry} and \eqref{eq:Lsymmetries}.

\begin{figure}
  \includegraphics{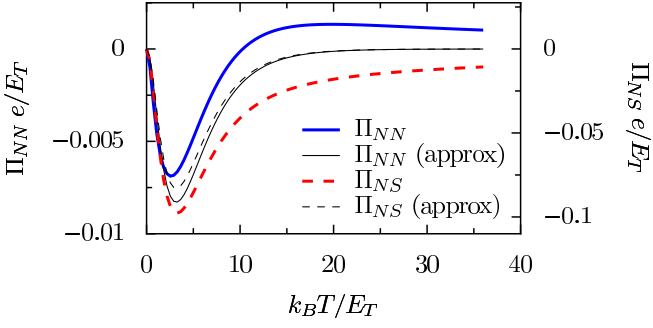}
  \caption{\label{fig:peltier} {(Color online)}
    Peltier coefficients $\Pi_{NN}$,
    $\Pi_{NS}$ for the same parameters as in Fig.~\ref{fig:Lmatrix}.
    Approximations \eqref{eq:peltierapprox},
    \eqref{eq:peltierNSapprox} are shown as thin lines --
    deviation from the exact result is due to neglecting $\T$.  }
\end{figure}

The finite coefficient $\tilde{L}_{LT}^{ij}(E)$ leads to a Peltier effect:
assume that the terminals are at a constant temperature $T^i =
\bar{T}$ and biased at potentials chosen so that a current $I_c$ flows
between terminals 1 and 2, $I_{c}^1 = -I_{c}^2 = I_c$.  Then, the
Peltier linear-response coefficient for this system is
\begin{equation}
  \Pi_{NN} \equiv \frac{dI_{Q}^1}{dI_c^1}
  = \frac{dI_{E}^1}{dI_c^1}
  =
  \frac{
    L_{21}^{11} W_1 - L_{21}^{12} W_2
  }{
    L_{11}^{11} W_1 - L_{11}^{12} W_2
  }
  \,,
\end{equation}
where $W_j \equiv (L_{11}^{1j} + L_{11}^{2j})^{-1}$.
We can also define the Peltier coefficient
$\Pi_{NS}\equiv\frac{1}{2}dI_{E}^1/dI_c^1$ corresponding to the current
configuration $I_c^1 = I_c^2 = I_c/2$.

The magnitude and temperature dependence of $\Pi$ is shown in
Fig.~\ref{fig:peltier}.  For a typical Thouless energy of
$E_T=\unit[200]{mK} \, k_B$ of an Andreev interferometer, the Peltier
coefficients would be $\abs{\Pi_{NN}}\sim\unit[100]{nV}$ and
$\abs{\Pi_{NS}}\sim\unit[1]{\mu{}V}$ at $T\sim\unit[200]{mK}$. For
comparison, Peltier coefficients for purely normal-metal junctions at
these temperatures are of the order $\Pi = T (S^B - S^A) \sim
\unit[0.2]{K} \times \unitfrac[10]{nV}{K}=\unit[2]{nV}$. The
interferometer induces a significantly larger $\Pi$.

The above Peltier effect is related to the thermopower discussed in
Refs.~\onlinecite{virtanen04,seviour00}. We indeed find the Kelvin
relations $\Pi_{NN} = T S_{NN}$, $\Pi_{NS} = T S_{NS}$, which follow
from the Onsager symmetry. Similarly as in
Ref.~\onlinecite{virtanen04b}, within the assumptions where
Eqs.~\eqref{eq:linearizedkin} apply, one can also find simple
approximations up to first order in $j_S$:
\begin{subequations}
\begin{align}
  \label{eq:peltierapprox}
  \Pi_{NN}
  &\approx\frac{(R_3-R_4)R_5^2}{2(R_1+R_2+R_5)(R_3 + R_4 + R_5)}
  \frac{k_B T}{e}
  \frac{\dd{I_{S,eq}}}{\dd{T}}
  \,,
  \\
  \label{eq:peltierNSapprox}
  \Pi_{NS}
  &\approx
  \frac{4R_3R_4R_5 + R_5^2(R_3+R_4)}{4(R_1+R_2+R_5)(R_3 + R_4 + R_5)}
  \frac{k_B T}{e}
  \frac{\dd{I_{S,eq}}}{\dd{T}}
  \,.
\end{align}
\end{subequations}
Here, $I_{S,eq} \equiv \frac{A\sigma}{2}\int_{-\infty}^{\infty}\dd{E}\,j_S\tanh\frac{E}{2k_BT}$ is
the equilibrium supercurrent. The above also shows the dependence on
the asymmetry for $\Pi_{NN}$ and the proportionality to the supercurrent
--- for this contribution to the effect.

\begin{figure}\centering
  \includegraphics{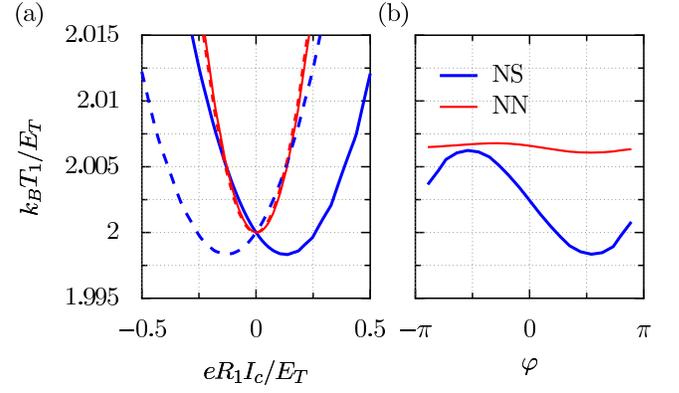}
  \caption{\label{fig:Tchange} {(Color online)} Temperature of
    terminal 1 in Fig.~\ref{fig:4probe} (a) for
    $\varphi=\arg\Delta_3-\arg\Delta_4 = \pm 0.54\pi$ (dashed and solid lines)
    and current configurations corresponding to $\Pi_{NN}$ and $\Pi_{NS}$
    (red and blue lines), as a function of $I_c$.
    (b) $T_1$ for $eR_1I_c/E_T=0.15$, as a function of $\varphi$.
    The results are
    calculated assuming other terminals are at the temperature
    $T=2E_T/k_B$. 
    Deviation of $T_1$ from $T$ originates from Joule heating
    and the oscillation of the proximity-Peltier effect.
  }
\end{figure}

Finite Peltier coefficients allow for cooling one of the terminals by
driving electric current. Assume the terminal is small enough, such
that the power flowing into the phonons is small compared to the heat
current carried by electrons. The temperature change is then limited
by the Joule heat generated in the wires: the heat current is $I_Q^1 =
-G_{th}\Delta T_1 -2 \Pi_{NS} I_c^1 + e(I_c^1)^2/G$, $G$ and $G_{th}$
being electrical and heat conductances. The maximum cooling effect
then is, in a rough estimate assuming that Wiedemann-Franz law
applies, $\Delta T_1 =
-(3/\pi^2)(e^2\Pi_{NS}^2/k_BT)/k_B\sim\unit[-0.3]{mK}$ for
$E_T=\unit[200]{mK}k_B$. Numerical calculation in the structure of
Fig.~\ref{fig:4probe} yields cooling $\Delta{}T\sim\unit[-0.4]{mK}$,
as shown in Fig.~\ref{fig:Tchange}.

One point to note is that also the $B$-symmetric oscillation of the
thermal conductance \cite{bezuglyi03,jiang04b} contributes to the
temperature change, although this is significant only at temperatures
small compared to $E_T/k_B$. In the absence of the Peltier effect,
$\Delta T$ would hence be symmetric in $B$ and always positive. The
proximity-Peltier effect allows negative temperature changes and also
breaks the symmetry, which makes the antisymmetric part
$T_1(B)-T_1(-B)$ the experimentally interesting signal. In the
structure of Fig.~\ref{fig:4probe} the oscillation amplitude can be
of the order of \unit[1]{mK} for $E_T=\unit[200]{mK}k_B$. (See
Fig.~\ref{fig:Tchange}b.) Temperature changes of this order can be
experimentally resolved in mesoscopic structures, \cite{meschke06} so
that the detection of the effect simply via observing $\Delta T$
should be experimentally viable.  In addition to the off-diagonal
thermoelectric coefficients $L_{12}$, $L_{21}$, it would also be
interesting to study the Onsager reciprocity for
$\tilde{L}^{ij}_{TT}(E)$ via differential conductances in
multi-terminal structures.

In summary, we have studied charge and energy transport and its
symmetry relations in normal--superconducting hybrid structures. We
show that a large Peltier effect controlled by the phase difference
over a Josephson junction can arise, partly due to co-flowing
quasiparticle and supercurrents.  This complements previous studies of
a related effect in the thermopower.


This research was supported by the Finnish Cultural Foundation
and the Academy of Finland.
We thank M.~Meschke and I.~A.~Sosnin for useful discussions.

\end{document}